\documentclass[showpacs]{revtex4}

\usepackage{graphicx}%
\usepackage{dcolumn}
\usepackage{amsmath}

\makeatletter
\def\btt#1{\texttt{\@backslashchar#1}}%
\DeclareRobustCommand\bblash{\btt{\@backslashchar}}%
\makeatother

\begin{document}

\title{Can Quintessence Be The Rolling Tachyon?}

\author{Xin-zhou Li}\email{kychz@shtu.edu.cn}

\author{Jian-gang Hao}
\author{Dao-jun Liu}
\affiliation{ Shanghai United Center for Astrophysics, Shanghai
Normal University,
Shanghai 200234 ,China\\
Institute for Theoretical Physics, ECUST, Shanghai 200237,China }

\date{\today}

\begin{abstract}

In light of the recent work by Sen\cite{Sen} and
Gibbons\cite{Gibbons}, we present a phase-plane analysis on the
cosmology containing a rolling tachyon field in a potential
resulted from string theory. We show that there is no stable point
on the phase-plane, which indicated that there is a coincidence
problem if one consider tachyon as a candidate of quintessence.
Furthermore, we also analyze the phase-plane of the cosmology
containing a rolling tachyon field for an exactly solvable toy
potential in which the critical point is stable. Therefore, it is
possible for rolling tachyon to be quintessence if one give up the
strict constraint on the potential or find a more appropriate
effective potential for the tachyon from M/string theory.
\end{abstract}

 \maketitle

\vspace{0.4cm} \textbf{1. Introduction} \vspace{0.4cm}

Recently, a field theory that describes the tachyon on a brane
-antibrane system near the minimum of its potential has been
proposed by Sen\cite{Sen}. Later on, Gibbons\cite{Gibbons}
considered the effective action of the tachyon coupling to
gravitational field on the brane by adding an Einstein-Hilbert
term. This further enticed the investigation of "tachyon
cosmology", which deal with the cosmology driven by the tachyon
rolling down to its ground state. Inflation as well as brane
cosmology driven by the rolling tachyon were also
studyed\cite{Fairbairn, Feinstein, Mukohyama, Gerasimov, Mazumdar,
Padmanabhan, Frolov, Minahan, Kim, Choudhury, shiu, Linde,
Maggiore, Sugimoto}. Especially, Kofman and Linde\cite{Linde}
pointed out the difficult for the rolling tachyon to drive the
inflation, i.e. it can not produce enough inflation that
compatible with the observation. So, it seems more suitable if we
consider the rolling tachyon as the dark energy that accelerates
the current expansion of the universe. It is worth noting that the
tachyon here is different from the supersymmetric tachyon we
investigated several years ago\cite{Li}.

On the other hand, the spectrum of CMB anisotropies have been
quite nicely fixed by current measurements \cite{Bernardis}, at
least for the initial region of small angles. Although these
observations should be extended to even larger red-shifts and
smaller angles, as several missions under preparation will do in
the near future, it is widely accepted that about 70 percent of
the total energy in the universe should be hidden as dark
energy\cite{Bahcall}. Observations of type Ia
supernova(SNIa)\cite{Perlmutter} shows that the expansion of
universe is accelerating and therefore requires that the
equation-of-state parameter $w=\frac{p}{\rho}$ of the total fluid
should be necessarily smaller than $-\frac{1}{3}$ , where $p$ and
$\rho$ are the pressure and energy density of the fluid in
universe respectively. Hitherto, two possible candidates for dark
energy have been suggested . One is the existence of a
cosmological constant and another is quintessence which is a
dynamical, slowly evolving, spatially inhomogeneous component
scalar field with negative pressure \cite{Ratra,Caldwell}. It is
widely accepted that a successful theory for dark energy must be
able to solve the "coincidence problem" and "fine-tuning
problem"\cite{Steinhardt}. Also, one would like to find a natural
quintessence model that can arise from high energy physics, or
eventually, resulted from string theory. So, it would be very
interesting to consider quintessence as the rolling tachyon of
string theory.

Now the question is that if tachyon field can be a candidate of
quintessence? Tachyon field holds negative pressure. So the key
point is that whether the rolling tachyon can solve the
"coincidence problem" and "fine-tuning problem". With the aid of
phase-plane analysis, we show that there is a saddle point on the
phase-plane, which indicates that the tachyon field need to be
fine tuned to account for the current cosmological observations.
It is worth noting that in the following discussion, like other
authors, we assume that there is no direct coupling between
tachyon and ordinary matter except through gravitational means.

\vspace{0.4cm}
\textbf{2. The Rolling Tachyon}
\vspace{0.4cm}

The effective Lagrangian density of tachyon of string theory in a
flat Robertson-Walker background is as following:
\begin{equation}
L=-V(T)\sqrt{1+g^{\mu\nu}\partial_\mu T\partial_\nu T}
\end{equation}

\noindent where
\begin{equation}
ds^{2}=-dt^{2}+a^{2}(t)(dx^2+dy^2+dz^2)
\end{equation}

is the flat Robertson-Walker metric and

\begin{equation}\label{VTk}
V(T)=V_0(1+\frac{T}{T_0})\exp(-\frac{T}{T_0})
\end{equation}

\noindent is the potential resulted from string
theory\cite{Kutasov}. It is not difficult to obtain the equation
of motion of the tachyon field as well as the gravitational field
as following:

\begin{equation}\label{H}
H^2=(\frac{\dot{a}}{a})^2=\frac{\kappa^2}{3}\rho_T
\end{equation}

\begin{equation}\label{dH}
\dot{H}=-\frac{\kappa^2}{2}(\rho_T + p_T)
\end{equation}

\begin{equation}\label{T}
\ddot{T}
+3H\dot{T}(1-\dot{T}^2)+\frac{V^{'}(T)}{V(T)}(1-\dot{T}^2)=0
\end{equation}

\noindent where the overdot represents the differentiation with
respect to $t$ and the prime denotes the differentiation with
respect to $T$. ${\kappa}^2=8\pi G$ where $G$ is Newtonian
gravitation constant. The density $\rho_T$ and the pressure $p_T$
are defined as following:

\begin{equation}
\rho_T=\frac{V(T)}{\sqrt{1-\dot{T}^2}}
\end{equation}

\begin{equation}
p_T=-V(T)\sqrt{1-\dot{T}^2}
\end{equation}

\noindent the equation-of-state parameter is

\begin{equation}
w=\frac{p_T}{\rho_T}=\dot{T}^2-1
\end{equation}

It is clear that if the tachyon field can accelerate the expansion
of the universe, there must be $\dot{T}^2<\frac{2}{3}$ and
$-1<w<-\frac{1}{3}$.

\vspace{0.4cm} \textbf{3. Phase-Plane analysis } \vspace{0.4cm}

Phase-plane analysis has been proved to be a very powerful tool
for us to investigate the behaviors of the field while not
necessarily solving the equation of motion. To obtain the
autonomous system corresponding to the equations of motion for
tachyon, we can substitute Eq.(\ref{H}) into Eq.(\ref{T}) and
obtain:

\begin{equation}\label{ddT}
\ddot{T}
+\sqrt{3}\kappa[\frac{V(T)}{\sqrt{1-\dot{T}^2}}]^{\frac{1}{2}}\dot{T}(1-\dot{T}^2)+\frac{V^{'}(T)}{V(T)}(1-\dot{T}^2)=0
\end{equation}

\noindent Now, introduce the new variables $X=T$ and $Y=\dot{T}$
and then the autonomous system is as following:

\begin{eqnarray}
\frac{dX}{dt}&=&Y \label{dX}\\
\frac{dY}{dt}&=&\frac{X(1-Y^2)}{T_0^2+T_0X}-\sqrt{3V_0}\kappa
Y(1-Y^2)^{3/4}(1+\frac{X}{T_0})^{1/2}\exp(-\frac{X}{2T_0})\label{dY}
\end{eqnarray}

It is not difficult to find the critical point of the phase-plane
as $(X,Y)=(0,0)$. Expand the above equations Eq.(\ref{dX}) and
Eq.(\ref{dY})about its critical point (0,0), we have

\begin{eqnarray}
\frac{dX}{dt}&=&Y \label{dX1}\\
\frac{dY}{dt}&=&\frac{X}{T_0^2}-\sqrt{3V_0}\kappa Y\label{dY1}
\end{eqnarray}

To determine the type of the critical points, we write down the
eigen-equation of the above system as:

\begin{equation}
\lambda^2+\alpha \lambda+\beta=0
\end{equation}

\noindent where $\alpha=\sqrt{3V_0}\kappa$ and
$\beta=-\frac{1}{T_0^2}$

\noindent One can easily found that the two eigenvalues are:

\begin{eqnarray}
\lambda_1&=&\frac{-\sqrt{3V_0}\kappa-\sqrt{3\kappa^2V_0^2+\frac{4}{T_0^2}}}{2} \label{lanmuda1}\\
\lambda_2&=&\frac{-\sqrt{3V_0}\kappa+\sqrt{3\kappa^2V_0^2+\frac{4}{T_0^2}}}{2}
\label{lanmuda2}
\end{eqnarray}

Clearly, we have

\begin{equation}
\lambda_1<0<\lambda_2
\end{equation}

This shows that the critical point of the autonomous system is a
saddle point and is unstable. That is to say, the evolution of the
tachyon field is very sensitive to the initial condition and  if
we consier tachyon field as quintessence, we must carefully
fine-tune it to match the current observations.

\vspace{0.4cm} \textbf{4. Phase-Plane Analysis for An Exactly
Solvable Toy Model} \vspace{0.4cm}

A toy model of tachyon has been introduced by
Feinstein\cite{Feinstein} to show how the so called power-law
inflation solution can be constructed. To do so, we must rewrite
the equations (Eq.(\ref{H}) to Eq.(\ref{T})) as:

\begin{equation}\label{dotT}
\dot{T}=-\frac{2}{3}\frac{H^{'}}{H(T)^2}
\end{equation}

\noindent and

\begin{equation}\label{H'}
 H^{'2}-\frac{9}{4}H^4(T)+\frac{1}{4}V^2(T)=0
\end{equation}

\noindent If one chooses the toy model potential

\begin{equation}\label{VT}
V(T)=AT^{\frac{4(n-1)}{2-n}}\sqrt{B+CT^{\frac{-2n}{2-n}}}
\end{equation}

\noindent where $A>0$,$B>0$ and $C<0$ are constants, and $0<n<1$
for the reality condition. From Eqs.(\ref{dotT}) and (\ref{H'}),
one has

\begin{equation}
a(t)=\exp(mt^n)
\end{equation}

\noindent and

\begin{equation}
T=\gamma t^{\frac{2-n}{2}}
\end{equation}

\noindent where $m$ and $\gamma$ are positive constants expressed
in terms of $A$, $B$ and $C$.

In the following, we will present a phase-plane analysis to show
that rolling tachyon with toy model potential Eq.(\ref{VT}). To do
so, we replace $V(T)$ in Eq.(\ref{ddT}) by Eq.(\ref{VT}), let
$x=T$ and $y=\dot{T}$ and rewrite Eq.(\ref{ddT}) into the form of
an autonomous system

\begin{eqnarray}
\frac{dx}{dt}&=&y \label{dx}\\
\frac{dy}{dt}&=&(1-y^2)\frac{4(n-1)B+(3n-4)Cx^{\frac{-2n}{2-n}}}{(2-n)x(B+Cx^{\frac{-2n}{2-n}})}-\sqrt{3A}\kappa
y(1-y^2)^{3/4}x^{\frac{2(n-1)}{2-n}}(B+Cx^{\frac{-2n}{2-n}})^{1/4}\label{dy}
\end{eqnarray}

It is easy to find that the only critical point of this system is
($2^{\frac{n-2}{n}}[\frac{B(1-n)}{C(3n-4)}]^{\frac{n-2}{2n}}$,0).
It is worth noting that judging the stability of the critical
point in general cases is a very complicated algebraic problem.
Here we only consider the special case for $n=\frac{2}{3}$. We
find that when $B>\frac{3}{64}A^2\kappa^4C^2$, the critical point
is a stable focus; when $B=\frac{3}{64}A^2\kappa^4C^2$, the
critical point is a stable degenerated node; and when
$B<\frac{3}{64}A^2\kappa^4C^2$, the critical point is a stable
node.

This shows that the critical point of the autonomous system is
always stable. That is to say, the evolution of the tachyon field
in this toy potential is not sensitive to its initial condition
and so there isn't so-called "coincidence problem" which exists in
the model with potential Eq.(\ref{VTk}).

Finally, let's briefly summarize the key aspects of this letter.
In this letter, we have investigated the possibility of
considering the rolling tachyon field as quintessence. We show, by
using the phase-plane analysis, that in the potential from string
theory there is no stable point in the phase-plane of the
tachyon's evolution equation. This indicates that if one considers
tachyon of string theory as the quintessence, he will encounter
the "coincidence problem". While, for the toy model we find that
there is a stable point to which the field will evolve. Although
the later one is only a toy model, it need not to particularly
choose the initial conditions of the tachyon field. In this
respect, the exactly solvable potential may be favorable.

\vspace{0.8cm}

This work was partially supported by National Nature Science
Foundation of China, National Doctor Foundation of China under
Grant No. 1999025110, and Foundation of Shanghai Development for
Science and Technology under Grant No.01JC14035.

\end{document}